\documentclass[aps, prl, twocolumn, superscriptaddress]{revtex4-1}
\usepackage{amsmath}
\usepackage{amssymb}
\usepackage{bm}   
\usepackage{verbatim} 

\def\({\left(} \def\){\right)}
\def\[{\left[} \def\]{\right]}
\def\al{\alpha} \def\bt{\beta}
\def\del{{\partial}}

\def\M{\mathcal{M}}

\newcommand{\non}{\nonumber \\}

\newcommand{\be}{\begin{equation}}
\newcommand{\ee}{\end{equation}}
\newcommand{\bea}{\begin{eqnarray}}
\newcommand{\eea}{\end{eqnarray}}
\newcommand{\ba}{\begin{eqnarray}}
\newcommand{\ea}{\end{eqnarray}}

\newcommand{\beq}{\begin{equation}}
\newcommand{\eeq}{\end{equation}}
\newcommand{\beqa}{\begin{eqnarray}}
\newcommand{\eeqa}{\end{eqnarray}}
\newcommand{\beqar}{\begin{eqnarray*}}
\newcommand{\eeqar}{\end{eqnarray*}}

\newcommand{\reef}[1]{(\ref{#1})}

\newcommand{\ie}{{\it i.e.,}\ }

\newcommand{\mt}[1]{\textrm{\tiny #1}}

\newcommand{\A}{\mathcal{A}}

\newcommand{\C}{\mathcal{C}}

\begin{document}

\title{Entanglement Entropy Flow and the Ward Identity}

\author{Vladimir Rosenhaus}
\email{vladr@berkeley.edu}

\author{ Michael Smolkin}
\email{smolkinm@berkeley.edu}

\affiliation{Center for Theoretical Physics and Department of Physics,\\
 University of California, Berkeley, CA 94720, U.S.A. }

\begin{abstract}
We derive differential equations for the flow of entanglement entropy as a function of the metric and the couplings of the theory.
The variation of the universal part of entanglement entropy under a local Weyl transformation is related to the variation under a local change in the couplings. We show that this relation is in fact equivalent to the trace Ward identity. As a concrete application of our formalism, we express the entanglement entropy for massive free fields as a two-point function of the energy-momentum tensor. 
\end{abstract}
\maketitle

\paragraph{\bf Introduction}

Entanglement entropy is a fundamental field theory quantity, having a broad range of applications \cite{Bomb86,Sred93,CalCardy04,RefMoore04,KitPres05,RT}. Yet, despite much recent progress, a general understanding of how to compute entanglement entropy, and how to relate it to more familiar field theory quantities, has remained limited. The goal of this note is to partially fill this gap. In particular, we derive equations encoding the change in the entanglement entropy under a variation of the metric and the couplings of the theory. We then use these equations to find a Ward-like identity involving the modular Hamiltonian.


\paragraph{\bf Ward Identity} Let us  consider  some QFT living on a Riemannian manifold $\mathcal{M}$ and assume that the action, $I$, contains a relevant or marginal operator $\mathcal{O}(x)$ of scaling dimension $\Delta\leq d$,
\be
\int_\M \lambda(x)\mathcal{O}(x)\subset I~,
\label{eq:action}
\ee
where we have promoted the coupling $\lambda$ to a background field $\lambda(x)$  \cite{foot0}. With this promotion, the action is now Weyl invariant provided that we scale $\lambda(x)$ appropriately \cite{foot1},
\be \label{eq:WeylTran}
 g_{\mu\nu}(x) \to e^{2\sigma(x)}g_{\mu\nu}(x)\, , \quad\quad \lambda(x)\to e^{(\Delta-d)\sigma(x)} \lambda(x) ~.
\ee 
Up to anomalies that are inherent to QFTs, the effective action is also invariant under (\ref{eq:WeylTran}), \ie it satisfies
\be \label{eq:Ward}
\left(2 g^{\mu \nu}(x){\delta \over \delta g^{\mu \nu}(x)} + (d- \Delta + \beta)\, \lambda(x)\, {\delta \over \delta \lambda(x)}\right) \Gamma_{eff} = -\mathcal{A} ~,
\ee
where $\mathcal{A}$ is the trace anomaly and $\beta$ is the anomalous dimension that emerges through renormalization in quantum field theory \cite{Nakayama:2013wda}. By definition, the various terms in (\ref{eq:Ward}) are given by one point correlation functions, and we get the celebrated trace Ward identity, valid to all orders in the coupling \cite{Osborn:1993cr,Jack:2013sha},
\be \label{eq:trace}
 \langle T(x)\rangle_{\lambda}+(d-\Delta+ \beta)\lambda(x)\langle \mathcal{O}(x) \rangle_{\lambda}=\mathcal{A}~.
\ee

Like the effective action, the entanglement entropy, $S_{\mt {EE}}(g_{\mu \nu}, \Sigma, \lambda)$, for an entangling surface $\Sigma$, is a fundamental quantity characterizing a field theory. In general, it consists of UV divergent local terms and a finite non-local contribution. Most of these terms depend on the choice of regularization scheme, however for an even dimensional $\M$ there is a logarithmic scheme-independent divergence.  The coefficient of this divergence, $S_{\text{univ}}$, is regarded as the universal part of entanglement entropy \cite{foot2}. 

Since $S_{\mt{EE}}$ is dimensionless, it is invariant under a constant rescaling of all dimensionful parameters (including the UV cut-off). Imposing this invariance on a general ansatz for $S_{\mt{EE}}$, it follows that the coefficient of the logarithmic divergence also shares this property. Thus, for constant coupling $\lambda$ 
\be
  \int_{\mathcal{M}} 2 g^{\mu\nu}(x) {\delta S_{\text{univ}} \over \delta g^{\mu\nu}(x)} + (d-\Delta+\beta)\,\lambda {\partial S_{\text{univ}}\over\partial \lambda}=0~.
  \label{eq:CS0}
\ee

Furthermore, for a CFT the universal part of $S_{\mt{EE}}$  is invariant under Weyl transformations \cite{foot3}. For a general QFT, by promoting the coupling constants to fields with the appropriate scaling, we restore Weyl invariance of $S_{\text{univ}}$, provided that the anomalous dimensions vanish. Therefore, it satisfies the analog of the Ward identity under (\ref{eq:WeylTran}),
\be  \label{eq:CS}
 \left(2 g^{\mu\nu}(x) {\delta  \over \delta g^{\mu\nu}(x)} + (d-\Delta+\beta)\,\lambda(x)\,{\delta \over\delta \lambda(x)}\right)S_{\text{univ}}=\mathcal{A}_{\mt{EE}}~ ,
\ee
where we have accounted for possible anomalies. We should note that there was no need to explicitly include variation of geometry associated with $\Sigma$, since rescaling of $g_{\mu\nu}$ induces rescaling of the extrinsic/intrinsic geometries of $\Sigma$.

Although we will not make use of it, in fact, since $S_{\mt{EE}}$ is invariant under a constant rescaling of all dimensionful parameters, it follows that the finite part of $S_{\mt{EE}}$ also satisfies a similar equation (\ref{eq:CS0}), the only difference being that for even dimensional $\mathcal{M}$ there is an anomalous piece on the right hand side. This anomaly is straightforwardly related to $S_{\text{univ}}$.


In what follows we derive equations satisfied by both terms in (\ref{eq:CS}):  equations describing how entanglement entropy changes under a variation of the metric (see Eq. \ref{eq:deltaSgeom}) and how it changes under a variation of the couplings (see Eq. \ref{eq:deltaSO1}). These two $\acute{a}$ priori  unrelated variations of the entanglement entropy are in turn linked by (\ref{eq:CS0}), (\ref{eq:CS}). We also derive the relation between (\ref{eq:CS}) and (\ref{eq:Ward}) \cite{foot44}. 

\paragraph{\bf Variation of Entanglement Entropy}
Consider some subregion $V$ of a manifold $\mathcal{M}$. The reduced density matrix for this region is obtained by tracing out degrees of freedom associated with $\overline{V}$ - the complement of $V$,
\be \label{eq:rho}
\rho = \text{Tr}_{\overline{V}} |0\rangle \langle 0| \equiv {e^{-K_\lambda}\over \text{Tr}_{\mt V} \, e^{-K_\lambda}},
\ee
where we have taken the global state to be the vacuum in the presence of $\lambda(x)$. The right hand side of (\ref{eq:rho}) serves as the definition of the modular Hamiltonian $K_\lambda$. The entanglement entropy is defined as the von Neumann entropy of the reduced density matrix, 
\be \label{eq:Sdef}
S_{\mt{EE}} = -\text{Tr}_V\,(\rho\log\rho)~.
\ee

Now let us consider any entangling surface $\Sigma$ and any given background configuration $\lambda_0(x)$ and $g_{\mu\nu}$. We are going to address the question of how  entanglement entropy changes under slight perturbations of the background configuration or the entangling surface. First, we evaluate the response of entanglement entropy to an infinitesimal change in the external field $\lambda_0 \rightarrow \lambda_0 + \delta \lambda$. Next, we find the change due to a slightly deformed metric $g^{\mu\nu}\rightarrow g^{\mu\nu} + \delta g^{\mu\nu}$ or a slightly deformed $\Sigma$.


To find these variations, we start with the Euclidean path integral representation of the reduced density matrix
\be \label{eq:rhoPath}
 [\rho]_{\phi_-\phi_+}={1\over \mathcal{N}}\int_{\phi(\C_+)=\phi_+  \above 0pt \phi(\C_-)=\phi_- } \mathcal{D}\phi \, e^{-I(\phi, g_{\mu\nu})}~,
\ee
where $\C_\pm$ are the two sides of a $(d-1)$-dimensional cut $\mathcal{C}$, such that $\del\,\mathcal{C}=\Sigma$, $\phi$ collectively denotes all the QFT fields,  and $\phi_{\pm}$ are some fixed field configurations on $\C_\pm$.

Consider now deforming the theory by a small change in the external field $\lambda_0(x)$. The resulting change in the reduced density matrix can be evaluated 
as a perturbative expansion in $\delta \lambda$ \cite{RS},
\be \label{eq:deltaRho}
[\delta \rho]_{\phi_-\phi_+}=-{1\over \mathcal{N}_0}\int_{\phi(\C_+)=\phi_+  \above 0pt \phi(\C_-)=\phi_- } \mathcal{D}\phi\int_{\mathcal{M}} \delta \lambda\, 
\big(\mathcal{O}-\langle \mathcal{O}\rangle_{\lambda_0} \big) \, e^{-I} + \ldots\, ,
\ee
where $\langle \cdots\rangle_{\lambda_0}$ denotes the connected vacuum expectation value in the unperturbed background. Hence, the  response to a small variation in the background field $\lambda_0(x)$ boils down to an integral over all possible insertions of $\mathcal{O}(x)$ into the path integral. The first order change in entanglement entropy is then found by appealing to the so-called first law of entanglement entropy, which follows from the first term in an expansion of (\ref{eq:Sdef})
\cite{Bhattacharya:2012mi,Blanco:2013joa,Wong:2013gua},
\be \label{eq:FirstLaw}
 \delta S_{\mt{EE}}= \text{Tr}(\delta \rho K_{\lambda_0}) ~,
\ee
where $K_{\lambda_0}$ is the modular Hamiltonian in the unperturbed background.
Now rewriting (\ref{eq:FirstLaw}) in terms of matrix elements, 
\be
\delta S_{\mt{EE}} = \int{d \phi_+ d \phi_- \langle \phi_+|\delta \rho| \phi_-\rangle \langle \phi_-| K_{\lambda_0} | \phi_+\rangle} ~,
\ee
and inserting (\ref{eq:deltaRho}), we get an unconstrained path integral over the entire manifold, which is just a correlation function in the vacuum state. Thus, the result for the first order change in the entanglement entropy is,
\be \label{eq:deltaSO}
\delta S_{\mt{EE}} = - \int_{\mathcal{M}}{\delta \lambda(x) \langle \mathcal{O}(x) K_{\lambda_0}\rangle_{\lambda_0}} +\ldots ~,
\ee
 Crucially, since this equation is valid for any $\lambda_0(x)$ and any manifold, one has the exact equation
\be \label{eq:deltaSO1}
{\delta S_{\mt{EE}} \over \delta \lambda(x)} = -\langle \mathcal{O}(x) K_{\lambda}\rangle_{\lambda} ~.
\ee
For the particular case of a constant field configuration, \ie a coupling constant, one can write (\ref{eq:deltaSO1}) in the form,
\be \label{eq:deltaSO1L}
 \frac{\partial S_{\mt{EE}}}{\partial \lambda} = -  \int_{\mathcal{M}}{\langle \mathcal{O}(x) K_{\lambda}}\rangle_{\lambda} ~.
\ee

Let us now keep the external field $\lambda_0(x)$ fixed and instead vary either the background metric $g^{\mu\nu}$ or the shape of the entangling surface $\Sigma$, or both. It turns out that through a proper choice of coordinates adopted to the entangling surface, both such deformations can be encoded in the variation of the metric, $\delta g^{\mu \nu}$.  The response to such deformations is \cite{RS} \cite{foot4},
\be
\delta S_{\mt{EE}} = -\frac{1}{2} \int_{\mathcal{M}}{\langle T_{\mu \nu}(x) K_{\lambda_0}\rangle_{\lambda_0} \delta g^{\mu \nu}(x) } +\ldots ~,
\label{varSmet}
\ee
where we have explicitly made use of the first law (\ref{eq:FirstLaw}), and the energy-momentum tensor is defined as
\be
 T_{\mu\nu} = \frac{2}{\sqrt{g}}\frac{\delta I}{\delta g^{\mu \nu}} ~.
\ee 
Since (\ref{varSmet}) is valid for any background field and metric, we obtain
\be \label{eq:deltaSgeom}
 \frac{\delta S_{\mt{EE}}}{\delta g^{\mu \nu}(x)} =-\frac{1}{2} \langle T_{\mu \nu}(x) K_{\lambda} \rangle_{\lambda} ~.
\ee

\paragraph{\bf Relation to the Ward Identity}
Eq. (\ref{eq:deltaSO1}) and (\ref{eq:deltaSgeom}) describe the response of entanglement entropy to local variations in the external field $\lambda(x)$ and geometry $(g^{\mu\nu},\Sigma)$, respectively. These two different kinds of variations are linked by (\ref{eq:CS0}),(\ref{eq:CS}). 
Similarly, the Ward identity (\ref{eq:Ward}) links the variations of the effective action with respect to geometry and couplings. 
It is interesting that we can, in fact, obtain (\ref{eq:CS0}),(\ref{eq:CS}) by use of  (\ref{eq:deltaSO1}), (\ref{eq:deltaSgeom}), and  (\ref{eq:Ward}).

We start with the simplest example: a flat background metric, $g_{\mu \nu} = \delta_{\mu \nu}$, and a planar entangling surface $\Sigma$. We denote the directions along $\Sigma$ by $y_i$ and the directions orthogonal to $\Sigma$ by $x_a$, such that $\Sigma$ is located at the origin $(x_1,x_2) = 0$. The region $V$ is thus the half-space,  $x_1>0$. With a  planar $\Sigma$, there is an $O(2)$ symmetry in the transverse space and the associated Killing field is  $\xi=x_1\del_2-x_2\del_1$. The generator of rotations is given by the analytic continuation of the Rindler Hamiltonian,
\be \label{eq:HR}
H_R = - \int_{\theta=\text{const}} {T_{\mu \nu} \xi^{\mu} n^{\nu} } ~,
\ee
where the polar angle $\theta$ in the transverse space to $\Sigma$ plays the role of Euclidean time, and $n^{\nu}$ is normal to a constant Euclidean time slice. As a result, the path integral (\ref{eq:rhoPath}) defining the reduced density matrix can be interpreted in terms of angular evolution of the state at $\theta=0$ to the state at $\theta=2\pi$, with the Rindler Hamiltonian being the generator of infinitesimal angular translations. This leads to the immediate conclusion that the modular and Rindler Hamiltonians are proportional \cite{KabStra},
\be \label{eq:Kplane}
K= 2\pi H_R ~.
\ee
We now take the trace Ward identity (\ref{eq:trace}) and differentiate with respect to $g^{\al\bt}(y)$ to obtain,
\bea  
 \langle T(x) T_{\al\bt}(y) \rangle_\lambda + (d-\Delta+\beta)\lambda(x)\langle \mathcal{O}(x) T_{\al\bt}(y) \rangle_\lambda 
 \non
 = {2\over\sqrt{g(y)}} \langle{\delta  T(x) \over \delta g^{\al\bt}(y)}\rangle_\lambda - {2\over\sqrt{g(y)}} {\delta\mathcal{A}\over\delta g^{\al\bt}(y)} 
\label{eq:trace2}
\eea
The right hand side reduces to a $\delta$-function with regular coefficient. 
In particular,  it does not contribute to $S_{\text{univ}}$ and we drop it in what follows.

Contracting (\ref{eq:trace2}) with $\xi^{\al} n^{\bt}$ and using (\ref{eq:HR}) and (\ref{eq:Kplane}) we obtain,
\be \label{eq:CS2}
 \langle T(x) K_{\lambda} \rangle_{\lambda} + (d - \Delta + \beta) \lambda(x) \langle \mathcal{O}(x) K_{\lambda}\rangle_{\lambda} =0~,
\ee
where we retained only terms that may contribute to the logarithmic divergence of $S_{\mt{EE}}$. Combining this identity with the flow equations (\ref{eq:deltaSO1}) and (\ref{eq:deltaSgeom}), we recover (\ref{eq:CS}) with $\mathcal{A}_{\mt{EE}}=0$.

For sure, our analysis so far strongly relied on the particular form of the modular Hamiltonian for a planar entangling surface. For a plane, $K_\lambda$ is given by an integral of the energy-momentum tensor, and therefore we could use the standard trace Ward identity to make the link between (\ref{eq:deltaSO1}) and (\ref{eq:deltaSgeom}), and obtain (\ref{eq:CS}). However, we now argue that for a general background $\mathcal{M}$ and a general entangling surface $\Sigma$, the flow equations (\ref{eq:deltaSO}) and (\ref{eq:deltaSgeom}) combined with the trace Ward identity yield (\ref{eq:CS0}),(\ref{eq:CS}). 

We start by rewriting (\ref{eq:trace}) on a conifold
\be  \label{eq:con}
 \langle T(x)\rangle_{\lambda}^\epsilon+(d-\Delta+ \beta)\lambda(x)\langle \mathcal{O}(x) \rangle_{\lambda}^\epsilon=\mathcal{A}_\epsilon~,
\ee
where $2\pi\epsilon$ is the deficit angle (or conical defect) that one introduces at the entangling surface to compute $S_{\mt{EE}}$ using the replica trick approach \cite{foot5,Callan:1994py,Fursaev:1994ea,Fursaev:1995ef}. Differentiating (\ref{eq:con}) with respect to $\epsilon$ and setting $\epsilon=0$ yields,
\be \label{eq:23}
  \del_\epsilon\langle T(x)\rangle_{\lambda}^\epsilon|_{\epsilon=0}+(d-\Delta+ \beta)\lambda(x) \del_\epsilon\langle \mathcal{O}(x) \rangle_{\lambda}^\epsilon|_{\epsilon=0}
  = \del_\epsilon\mathcal{A}_\epsilon|_{\epsilon=0}~.
\ee
Now using the result from  \cite{SS} that differentiation with respect to the conical deficit corresponds to insertion of the modular Hamiltonian, the left hand side of (\ref{eq:23}) can be expressed as,
\bea
  \langle T(x) K_{\lambda} \rangle_{\lambda} + (d - \Delta + \beta) \lambda(x) \langle \mathcal{O}(x) K_{\lambda}\rangle_{\lambda}
  \non
  = \del_\epsilon\mathcal{A}_\epsilon|_{\epsilon=0}+\langle \del_\epsilon T(x)\rangle_{\lambda}|_{\epsilon=0}~,
  \label{eq:deficit}
\eea
where the last term stands to emphasize that the energy-momentum tensor is sensitive to the deficit angle as it explicitly depends on the metric. Setting $\lambda(x)$ to a constant and integrating (\ref{eq:deficit}) over $\mathcal{M}$ yields (\ref{eq:CS0}). The integral of the anomalous term on the right hand side of (\ref{eq:deficit}) is a finite term localized on the entangling surface and therefore does not contribute to the logarithmic divergence of $S_{\mt{EE}}$ \cite{foot6}. Consequently, we are justified in dropping the right hand side of (\ref{eq:deficit}), thus recovering (\ref{eq:CS2}) and hence (\ref{eq:CS}). In fact, this analysis indicates that $\mathcal{A}_{\mt{EE}}=0$ in general. However, this conclusion rests on the conifold method \cite{Fursaev:1995ef,Fursaev:1994ea,Lewkowycz:2013nqa,Fursaev:2013fta} which is not understood in full generality. It would be interesting to gain a better understanding of $\mathcal{A}_{\mt{EE}}$.



\paragraph{\bf Free Fields}
As a simple illustration of our findings, we evaluate $S_{\text{univ}}$ for a planar entangling surface $\Sigma$ embedded in flat space for a free Dirac field of mass $m$. In this case $\lambda(x)=m$, and the flow equation (\ref{eq:deltaSO1L}) takes the  form
\be
  {\del \over\del m} S_{\text{univ}}=2 \pi \int d^d x \int_{\Sigma}\int_0^\infty d\bar{x}_1\, \bar{x}_1 \langle \, O(x) \, T_{22}(\bar{x}_1,0,\bar{y}_i)\,\rangle_m ~, 
\ee
where we used (\ref{eq:Kplane}) and chose $\theta=0$.
Alternatively, one can make use of (\ref{eq:CS}) combined with (\ref{eq:deltaSgeom}) to write
\be \label{eq:Sferm2}
  {\del \over\del m}S_{\text{univ}}=-{2\pi\over m} \int d^d x \int_{\Sigma}\int_0^\infty d\bar{x}_1\ \bar{x}_1 \langle \, T(x) \, T_{22}(\bar{x}_1,0,\bar{y}_i)\rangle_m ~,
\ee
where the scaling dimension of the fermionic mass operator is $\Delta = d-1$ and $\beta = 0$.

As opposed to the standard representation of entanglement entropy in terms of heat kernels, here we have expressed it in terms of a $2$-point function. Evaluating the integrals in (\ref{eq:Sferm2}) yields \cite{RS3} 
\be
 S_{\text{univ}}={(-)^{d/2} \over 6(2\pi)^{d-2\over 2}\Gamma(d/2)} m^{d-2}\A_\Sigma \log(m \delta) ~.
\ee
where by assumption $d$ is even and $\delta$ is the UV cut-off. Thus we have recovered the universal `area' term for fermions \cite{KabStra,Hertzberg:2010uv,Huerta:2011qi,Lewkowycz:2012qr}.


\paragraph{\bf Discussion}
In this note we have derived equations for the flow of entanglement entropy in the space of QFTs: flow both in the couplings of the operators (\ref{eq:deltaSO1}), and in the geometry (\ref{eq:deltaSgeom}). By promoting the couplings to fields, the Weyl invariance of the resulting `ambient' action led to a nontrivial link (\ref{eq:CS}) between these two very different kinds of variations of the theory. This relation is  in the spirit of the Ward identities: relations among correlation functions of local operators that follow from a symmetry of the theory. Our relation, however, is among correlation functions involving the modular Hamiltonian, a quantity which is generally nonlocal.

Entanglement entropy is a relatively new concept within field theory. It is a nonlocal quantity, one which is nontrivial to measure \cite{KlichLev08}, and one whose computation is not easily susceptible to standard QFT techniques. Therefore, expressing entanglement entropy in terms of correlators of local operators is an essential ingredient in the entanglement entropy dictionary. For certain symmetric entangling surfaces, such as a plane, the modular Hamiltonian is local, and our flow equations provide this entry. 

While mysterious as a field theory quantity, entanglement entropy is, completely remarkably, given holographically by an area of an extremal surface \cite{RT}. Perhaps then, trying to reduce entanglement entropy to correlation functions is outdated. Rather, we should regard entanglement entropy as the basic building block and Eqs. (\ref{eq:CS}), \reef{eq:deltaSO1}, \reef{eq:deltaSgeom} as a way to recover the standard field theory quantities. 

\paragraph{Acknowledgments}
We thank H.~Casini, L-Y.~Hung, R.~Myers, M.~Rangamani, S.J.~Rey, and S.~Solodukhin for helpful discussions. This work is supported in part by NSF Grant PHY-1214644, and by the Berkeley Center for Theoretical Physics.

\begin{thebibliography}{999}
\bibitem{foot1}{ To maintain Weyl invariance of the full action, we promote every dimensionful coupling to a field and require that it transforms appropriately under a local rescaling of the metric. As a result, one has to sum over all such fields in \reef{eq:Ward}. Without loss of generality we assume that $\mathcal{O}(x)$ forms a basis for the trace of the energy momentum tensor and the theory is conformal when $\lambda(x)$ vanishes.}
\bibitem{foot4}{In \cite{RS} the focus was on background geometries such as a flat plane in Euclidean space. These symmetric setups allowed evaluating the path integral expression for $\delta \rho$. Here we are considering a general manifold for which this is not possible. However, this has no impact on the validity of our final equation for $S_{\mt{EE}}$.}
\bibitem{foot5}{To avoid singularities one can carry out computations on a regularized cone and remove all regulators at the very end.}
\end{thebibliography}

\end{document}